\documentclass[12pt]{article}
\usepackage[margin=0.9in]{geometry}
\usepackage{graphicx} 
\usepackage{amsmath}
\usepackage{subfigure}
\usepackage{widetext}
\providecommand{\keywords}[1]{\textbf{Keywords:} #1}

\date{}	
	\title{Evolution of fluctuations in horizon energy and its dependence on the degrees of freedom}
	\author{ Vishnu S Namboothiri$^{1},$ P.B. Krishna$^{2},$ Adithya P.S.$^{3},$ Titus K. Mathew$^{4}$\\[0.1in]
              Department of Physics, Cochin University of Science and Technology, \\ Kochi-22, India. \\[0.08in]
              vishnus@cusat.ac.in$^{1},$ krishnapb@cusat.ac.in$^{2},$ \\ adithyaps@pg.cusat.ac.in$^{3},$ titus@cusat.ac.in$^{4}$}
              
\begin{document}
              
              \maketitle 
 \begin{abstract}
	Taking account of the thermal nature of the Hubble horizon of the expanding universe, we analysed the evolution of relative fluctuations of horizon energy. For this analysis, we used two approaches: (i) by treating the Hubble horizon as a system in canonical ensemble, and (ii) by considering the microscopic degrees of freedom on the horizon. In both approaches, we obtained the relative fluctuations by using two different definitions of the horizon temperature; first, the Gibbons-Hawking temperature, and second, the Kodama-Hayward temperature. For a given temperature, both approaches yield the same general evolution for the fluctuations. In the asymptotic limit, the relative energy fluctuations corresponding to the Gibbons-Hawking temperature, is  $[{\hbar G}/{2\pi}] H^2,$ and $2/N_{sur}$ for the first and second approaches respectively. 
	Similarly, using the Kodama-Hayward temperature, the asymptotic fluctuations are $[{5\hbar G}/{2\pi}] H^2,$ and $10/N_{sur}.$ 
	This implies that, the magnitude of the relative fluctuations of the horizon energy is higher in the case of Kodama-Hayward temperature.
	The inverse dependence of the fluctuation on $N_{sur},$ the number of degrees of freedom on the horizon, reflects a familiar behaviour in ordinary thermal systems: fluctuations decrease as the number of degrees of freedom increases. Notably, we also found that the relative energy fluctuations establish a connection between the Planck length scale $L_p,$ characteristic length scale of the very early epoch of the universe, and $\sqrt{3/\Lambda},$ the length scale associated with the late-time accelerated phase. This relationship can offer valuable insights that could help in addressing the cosmological constant problem.

\end{abstract}

\section{Introduction} \label{section 1}
The thermodynamic nature of black hole horizon is a well established fact\cite{Hawking:1976de,Bardeen:1973gs}. 
Hawking had shown that black holes can emit radiations just like a black body\cite{Hawking:1975vcx,Gibbons:1977mu}, with temperature proportional to its surface gravity.  Meantime, Bekenstein have shown that Black hole horizon do possess entropy too, which is proportional to its horizon area\cite{Bekenstein:1973ur,Bekenstein:1975tw}. Thermodynamic nature is not a unique property of the black horizons, but was generalised to other spacetime horizons too. An important generalisation is in the context of cosmology. It was established convincingly that, like black holes, the horizon of an expanding universe possess temperature proportional to the Hubble parameter\cite{Gibbons:1977mu} and hence have entropy proportional to the horizon area\cite{Bousso:1999xy,Bousso:2002ju}. More interestingly, it has been shown that, the apparent horizon or the Hubble horizon of the flat universe is associated with  entropy, $S_H = A/(4G),$ and temperature $T=\kappa/(2\pi),$ where $A$ is the horizon area and $\kappa$ is the surface gravity at the Hubble horizon of the FLRW universe\cite{Bousso:2004tv}. Thus horizons in general are thermodynamic in nature, which was further supported by the discovery of Unruh and Davies, regarding the thermal property of the Rindler horizon experienced by an accelerator observer in flat spacetime\cite{Unruh:1976db,Davies:1974th}. 

An inherent property of any thermal system is the presence of energy fluctuations, arising from the random or statistical nature of the associated heat energy. Given the thermodynamic character of cosmic horizons, it is natural to expect that the energy associated with the cosmological horizon will also exhibit thermal fluctuations, which evolve as the universe expands. Several studies in the literature have analysed thermal fluctuations of the matter fields in the universe to investigate their potential contribution to the generation of cosmic structures.   The primordial thermal fluctuations associated with the cosmic components has been considered in references\cite{Biswas:2013lna,Magueijo:2002pg,graef2021constraining}. There are other works too which investigate the fluctuation of energy of the cosmic components within the horizon of the universe, for details see the references \cite{Magueijo:2007wf,Magueijo:2006fu,Biswas:2014kva,Brandenberger:2006vv,brandenberger2009string,Li:2015egy,Cai:2009rd}. All of the above works are analysing the fluctuations in the volume energy $\rho V$ (where $\rho$ is the density of the cosmic matter and $V$ is the volume enclosed by the horizon) within the horizon, in the context of early universe, and mainly tries to explain the primordial fluctuations, which causes the structure formation in the universe\cite{Peebles:1982ff,Peebles:1994xt,rephaeli1986correlated}. But only little has been explored about the evolution of thermal fluctuations of the horizon energy itself. Mimoso and Pavon have studied the fluctuations of the energy flux across the apparent horizon of a homogeneous and isotropic universe\cite{Mimoso:2018juu}. Very recently, the fluctuations in the energy, stored on the Hubble horizon of the end epoch of the universe, the de Sitter phase, have been analysed by Komatsu\cite{Komatsu:2021ncs}, by assuming Hawking temperature for the horizon. However, there remains a significant gap in the literature regarding the analysis of the evolution of energy fluctuations associated with the cosmic horizon. 

Investigating the evolutions of fluctuations can yield valuable insights into the stability of the horizon itself. Further, understanding the evolutions of the thermodynamic fluctuations of horizon energy is particularly significant, as it may shed new light on the nature of dark energy and potentially offer clues toward resolving the cosmological constant problem
\cite{B:2019oil,Krishna:2017vmw}. In this work, we study the evolution of thermal fluctuations in the energy of the Hubble horizon using two approaches: first, through the framework of the canonical ensemble, and second, by employing the degrees of freedom associated with the horizon. 

In the first approach, which follows the conventional method of analysing fluctuations in any thermal system, the canonical ensemble approach, we define a partition function for the horizon as the sum of the Boltzmann factor, $e^{-\beta E_r},$ over all possible energy states of the horizon. Using the partition function, the evolution of the fluctuations can be analysed in a standard way. The second approach is a novel one, based on the degrees of freedom on the Hubble horizon. Any thermal system possess microscopic degrees of freedom to carry the heat energy. For ordinary matter, the conventional method to obtain this degrees of freedom is through the relation $E/[(1/2)k_{_B} T],$ where $E$ is the energy of the system, $T$ is the temperature and $k_{_B}$ is the Boltzmann constant. But extending this formula to gravitating systems, especially to cosmological horizon is not straight forward.
The question is then, how to obtain the degrees of freedom residing on the horizon. For this we follow the conjuncture proposed by Padmanabhan, which gives a well defined method for it. We will briefly describe the time line of the emergence of this conjuncture. 
Following the pioneering work of Hawking and Bekenstein, which first revealed the intriguing relationship between gravity and thermodynamics, Jacobson  derived Einstein’s field equations from the Clausius relation by applying it to a local Rindler causal horizon\cite{Jacobson:1995ab,Eling:2006aw}.  Meanwhile, Padmanabhan derived Newton's law of gravity by combining the equipartition law of energy and the thermodynamic principle 
$S=E/2T$ on the horizon for a suitably defined accelerating observer, where  $S$ is the horizon entropy, $T$ is the horizon temperature, and 
$E$ is the active gravitational mass within the horizon\cite{Padmanabhan:2003pk,Padmanabhan:2009kr,Padmanabhan:2012bs}. This connection was further verified in the context of higher-order gravity theories, such as Gauss-Bonnet and Lovelock theories\cite{Cai:2005ra,Akbar:2006kj,Akbar:2006er,Hareesh,Mahith,VT,PhysRevD.107.063511}. This had led many to propose that gravity could be an emergent phenomenon similar to thermodynamics\cite{Padmanabhan:2004kf}. There are attempts to extend the concept of emergence, so as to consider the spacetime itself as emergent\cite{Padmanabhan:2008zza,Padmanabhan:2007tm}. However, it is difficult to treat time as being emerged from a more fundamental object. But there is a solution to this difficulty in cosmology, where all fundamental co-moving observers who are relatively at rest with respect to each other and for whom the cosmic microwave background radiation is homogeneous and isotropic,  measures the same time, their proper time, known as the cosmic time. Due to this it is possible to bifurcate time from cosmic space. This enable one to postulate that the cosmic space is emerging with the progress of cosmic time. Following this, Padmanabhan have conjectured that, the time evolution of the volume of the Hubble sphere (equivalent to the expansion of universe) is driven by the holographic discrepancy, $N_{surf} - N_{bulk},$  between the degrees of freedom, 
where $N_{surf}$ is the degrees of freedom of space on the surface of the Hubble horizon and $N_{bulk}$ is the degrees of freedom of the gravitating matter that has been emerged within the volume bounded by the horizon\cite{padmanabhan2012emergence,Krishna:2017vmw}. Here the horizon degrees of freedom is proposed to be proportional to area of the horizon. Assuming, $L_p^2,$ the Planck area, corresponding to one degree of freedom, it is possible to obtain the horizon degrees of freedom as the ratio of the horizon area with Planck area. 
The fluctuations of energy in any thermal system is strongly related to the degrees of freedom residing in the system. Same can be expected to be true in the case of Hubble horizon also. We will calculate the fluctuation of the horizon energy by obtaining the energy of the horizon using the equipartition rule, which is related to its degrees of freedom. 

Temperature of the horizon is an important parameter required for obtaining the energy fluctuations. Gibbons and Hawking \cite{Gibbons:1977mu} have defined the temperature of the cosmological horizon as $T=\hbar H/2\pi k_{_B},$ which has been used by many as the temperature of the Hubble horizon in various contexts. We first used this temperature to study  the evolution of the fluctuations. However, it was later argued that the temperature of the horizon in an expanding universe should also depend on the time evolution of the Hubble parameter. Incorporating this dynamical aspect, Kodama and Hayward proposed a modified expression for the horizon temperature\cite{Muhsinath:2022cij}, now known as the Kodama–Hayward temperature. We then obtain the expression for the fluctuations in horizon energy by using the Kodama-Hayward temperature.
Further, we analyse the thermal fluctuations in the energy of the Hubble horizon of a homogeneous and isotropic universe 
in the context of the standard $\Lambda$CDM model. 

Furthermore, we explore the possibility of addressing the cosmological constant problem. In a previous work, the cosmological constant is obtained as an integration constant, by integrating the equation that gives the time rate of change of the Hubble parameter\cite{Gob}. It is worth noting that, in the present work, we by making use of the dependence of the relative energy fluctuations on the degrees of freedom on the horizon, analyse the possibility of extracting the magnitude of cosmological constant. The motivation for this basically arises from the dependence of the surface degrees of freedom on both the Planck length and the Hubble parameter. We will show that, if an independent method to measure the relative energy fluctuations of the horizon energy can be established, then it would offer a novel approach to determine the magnitude of the cosmological constant and can potentially lead to a resolution of the cosmological constant problem.

The article is arranged in the following way. In section II we analyse the evolution of relative energy fluctuations on the horizon following the canonical ensemble approach by considering the Gibbons-Hawking temperature and Kodama-Hayward temperature for the horizon. We also include in the same section, our analysis of the evolution of the relative fluctuations in the context of the $\Lambda$CDM model. In section 3, we have present a similar analysis with Kodama-Hayward temperature for the horizon. In the last section, we present the conclusions. It may be noted that throughout our analysis we took, the speed of light, $c=1.$

\section{Horizon Energy fluctuations - the canonical formalism}

The fluctuations in energy can be occurred in a thermodynamic system with a finite temperature, which arises from the random motion of microscopic constituents, such as atoms or molecules, within the system. In this paper, we apply conventional statistical methods to analyse the horizon energy fluctuations. However, unlike in the traditional Gibbsian mechanics, the entropy of the Hubble horizon (Bekenstein entropy) is non-extensive, being proportional to the horizon's area rather than its volume. Notably, Hawking, in his analysis of black hole radiation, employed the microcanonical ensemble approach - an integral part of Gibbsian statistical mechanics, utilising Bekenstein entropy\cite{Gibbons:1977mu,Wald:1999vt}. To obtain energy fluctuations, we adopt the canonical ensemble approach, as the energy in a microcanonical ensemble remains constant. Previous studies have highlighted that applying the canonical ensemble approach to highly gravitating systems, such as black holes, leads to divergence in the partition function due to their negative heat capacities\cite{Lyn,Thir,Vel}. However, in a cosmological context, the situation differs. The observable universe consists of a bulk region bounded by the cosmological horizon, and if the magnitude of the horizon's heat capacity (which is negative) exceeds that of the bulk, then the system remains stable, making the canonical ensemble a viable tool for studying its thermodynamic properties.
Many results in horizon thermodynamics are based on Gibbsian statistical mechanics \cite{Gibbons:1977mu,Wald:1999vt}. Given the complexity of the issue, employing the canonical ensemble as a working assumption remains a practical approach. This approach allows for a comprehensive understanding about the contribution of the random movements of the particles to the overall variability in thermodynamic properties.  
The partition function of a canonical system  is given by,      
\begin{equation}
Z(\beta)=\sum_{r} e^{-\beta E_{r}}.
\end{equation}
The summation spans across all possible
energies, $E_r,$ that the system can possess, while the parameter $\beta=(1/k_{_B} T),$ where $k_{_B}$ represents the Boltzmann constant and $T$ is the temperature of the system. Now, the total energy, $U$ of the thermodynamic system, equivalent to the ensemble average $\left<E\right>,$ can be 
expressed as,
\begin{equation}
U=\left\langle {E} \right\rangle= \frac{\sum_{r}E_{r}e^{-\beta E_{r}}}{\sum_{r} e^{-\beta E_{r}}} = -\frac{\partial \ln Z}{\partial \beta}.
\end{equation} 
The fluctuations in energy can then be obtained as 
\begin{equation}\label{eqn:fluc123}
\sigma^2_{E}=\langle E^2 \rangle-\langle E \rangle^2 
= \frac{\partial^2 \log Z}{\partial\beta^2}=-\frac{\partial U}{\partial \beta}=k_{_B}T^2c_{_V},
\end{equation} 
where,
\begin{equation}
\left\langle E^2 \right\rangle=\left(\frac{1}{Z}\frac{\partial^2 Z}{\partial \beta^2}\right),
\end{equation}
is the average of the square of the energy, and 
$c_{_V}$  is the heat capacity of the system for a given volume. The above results reveal that, the fluctuations of the 
energy in any system is in general proportional to the heat capacity 
and also to the square of the temperature of the system\cite{chatterjee303030black,Artymowski:2018pyg}. We will follow this basic formalism to obtain the fluctuations in the horizon energy\cite{Landau:1980mil}. 

\subsection{\textbf{Fluctuations in horizon energy - Assuming the Gibbons-Hawking relation for temperature}}

Here, we are 
studying thermal fluctuations of horizon energy 
using the Gibbons-Hawking temperature. This temperature is initially proposed as a measure of the temperature of the de Sitter horizon of universe. It is expressed as\cite{Gibbons:1977mu},
\begin{equation}
T_H = \frac{\hbar H }{2\pi k_{_B}}.
\end{equation}
This same expression was used to represent the temperature of the  dynamical apparent horizon of the expanding universe\cite{Gibbons:1977mu,Muhsinath:2022cij}. Since the Hubble parameter $H$ is basically a dynamical quantity for an expanding universe, it was argued that, this form of the temperature is relevant in extracting the thermodynamics of the dynamical horizon.

To obtain the fluctuations in energy of the Hubble horizon of the expanding universe, we first need to have the expression for the horizon energy.
The active gravitational energy of the horizon,  
is related to its thermodynamic properties, the entropy $S_{_H},$ and temperature, $T_{_H}.$ As shown and discussed in detail in reference\cite{Padmanabhan:2012bs}, the average energy of the horizon is given by, 
\begin{equation}\label{eqn:HE1}
\left<E_{_H} \right> = 2 T_{_H} S_{_H}.
\end{equation}
According to Bekenstein\cite{Bardeen:1973gs,Bekenstein:1973ur,Srednicki:1993im}, the entropy of the horizon is proportional to  
the area of the horizon in the context of a black hole. This idea was extended to cosmology by Hawking and others\cite{Gibbons:1977mu,Bousso:1999xy,easther1999holography}. Accordingly, the entropy of the Hubble horizon of a flat  FLRW universe is given by,
\begin{equation} \label{Hentro1}
S_{_H} = \left(\frac{k_{_B}}{\hbar G} \right) \frac{A_{_H}}{4},
\end{equation}
where 
$G$ is Newton's gravitational constant, $\hbar$ is the reduced Planck constant, and $A_H=4\pi r_{_H}^2$ is the area of the Hubble horizon of the universe, with radius $r_{_H}=1/H.$ The energy of the Hubble horizon can be calculated using equation(\ref{eqn:HE1}) as
\begin{equation}
E_{_H}=\frac{1}{GH}.
\end{equation}
Having the expression for energy, we will then obtain the fluctuations of horizon energy, following the previous section as,
\begin{equation}
\sigma^2_{_E}=\frac{\hbar}{2\pi G}.
\end{equation}
The relative energy fluctuations on the Hubble horizon can then be obtained as,
\begin{equation}
\label{eqn:flucdof12}
\left|\frac{\sigma^2_{_E}}{\left\langle {E_H} \right\rangle^2}\right|=\frac{\hbar G}{2\pi}H^2.
\end{equation}
Since the Hubble parameter decreases with the expansion of the universe, the relative fluctuations in horizon energy also seems to decrease as per the above relation. 
This decrease in the fluctuations will continue until the universe attain the end de Sitter epoch, at which the relative energy fluctuations becomes a constant minimum value.

\subsection{
	\textbf{Fluctuations in horizon energy - Assuming the Kodama-Hayward relation for temperature}}
Now, we analyse the evolution of the fluctuations of the energy of the Hubble horizon using the Kodama-Hayward temperature.  
To obtain the fluctuations in energy of the Hubble horizon of the expanding universe, we first need to have the expression for the horizon energy.
The active gravitational energy of the horizon,  
is related to its thermodynamic properties, the entropy $S_{_H},$ and temperature, $T_{_H}.$ As shown and discussed in detail in reference\cite{Padmanabhan:2012bs}, the average energy of the horizon can be obtained using equation (\ref{eqn:HE1}).

The entropy of the Hubble horizon of a flat  FLRW universe is given in equation(\ref{Hentro1}). 
The temperature of the horizon is proportional to its surface gravity\cite{Bousso:2004tv,Cai:2008gw},
\begin{equation}\label{eqn:kappa}
\kappa = -H \left(1+ \frac{\dot H}{2H^2} \right).
\end{equation}\label{eqn:T2}
Consequently, the temperature is then given by 
\begin{equation}\label{eqn:T3}
T_H = \frac{\hbar \left|\kappa \right|}{2\pi k_{_B}}.
\end{equation}
In defining the temperature, the modulus of $\kappa$ is taken for having the horizon temperature positive, which implies a positive heat capacity for the universe. Consequently, the universe will be in a thermodynamically stable state\cite{hashemi2015hawking,Sadeghi:2014gza}. 
Here we have used the Kodama-Hayward temperature because 
it contains terms proportional to the time derivative of the Hubble parameter. It is widely accepted that throughout the evolution of the universe, equilibrium or at least quasi-static equilibrium is maintained and hence we can apply equilibrium thermodynamics to study the evolution. Numerous studies support the argument that the slow rate
of expansion allows the universe to remain in local thermodynamic equilibrium. 
In this sense, various works on deriving the Einstein field equations from thermodynamics rely on the assumption of local thermodynamic equilibrium in an expanding universe\cite{Banihashemi:2022htw, Banihashemi:2022jys}.  
Using equations (\ref{Hentro1}) and (\ref{eqn:T3}), the horizon energy given in (\ref{eqn:HE1}), can be expressed as, 
\begin{equation}
\left< E_H \right> =\frac{1}{G}\frac{1}{H}\left(1+\frac{\dot H}{2H^2}\right).
\end{equation}
Having this expression for energy, the horizon energy fluctuations can be obtained following the procedure described in the previous section. In this context, we assume that the horizon of the universe is a system of canonical ensemble. This is a viable assumption, because the fluid in the universe is a thermal fluid in significant abundance and hence the interactions within the fluid are of scales more prominent than the Hubble expansion rate, which is adequate to maintain a local thermal equilibrium \cite{Mazumdar:2010sa,Allahverdi:2010xz}.  
Accordingly, we arrived at the expression for the energy fluctuations at the Hubble horizon, with the help of equation (\ref{eqn:fluc123}),
\begin{equation}\label{eqn:sigma2}
\sigma_{_H}^2 
=\frac{1}{2\pi}\frac{\hbar}{G}H^2\left(1+\frac{\dot H}{2H^2}\right)^2\frac{\left(-\frac{1}{H^2}-\frac{3}{2}\frac{\dot H}{H^4}+ \frac{d \dot H}{d H}\frac{1}{2H^3}\right)}{\left(1-\frac{\dot H}{2 H^2}+\frac{d\dot H}{dH}\frac{1}{2H}\right)}.
\end{equation} 
From this, the relative fluctuations in energy at the Hubble horizon can be obtained as, 
\begin{equation}\label{eqn:fluc11}
\frac{\sigma_{_H}^2}{\left<E_{_H}\right>^2}=\frac{\hbar  G}{2\pi}H^2\frac{\left(-1-\frac{3}{2}\frac{\dot H}{H^2}+ \frac{d \dot H}{d H}\frac{1}{2H}\right)}{\left(1-\frac{\dot H}{2 H^2}+\frac{d\dot H}{dH}\frac{1}{2H}\right)}.
\end{equation}
According to this, the relative energy fluctuations will evolve with the expansion of the universe and 
is depending on the Hubble parameter, its time rate, and the rate of $\dot H,$ with respect to the Hubble parameter itself.

\subsection{\textbf{Evolution of the horizon energy fluctuations in the context of $\Lambda$CDM model}}

Here, we aim to obtain the nature of the evolution of the fluctuations of horizon energy, in the context of the standard $\Lambda$CDM model. It has to be highlighted that, the final expressions of the relative fluctuations corresponding to the two temperatures are different. We will look at this difference, by considering the equations (\ref{eqn:flucdof12}) and (\ref{eqn:fluc11})
in the context of the $\Lambda$CDM model.

In $\Lambda$CDM model the Hubble parameter, as a function of the scale factor of expansion, is given by,
\begin{equation}
H^2=H_{_0}^2\left(\Omega_{_{m0}}a^{-3}+\Omega_{_{\Lambda}}\right),
\label{eqn:Hubble1}
\end{equation}
where $H_{_0}$ is the present value of the Hubble parameter, $\Omega_{_{m0}}$ is the present mass density parameter of non-relativistic matter, $\Omega_{_{\Lambda}}$ is the density parameter of cosmological constant and $a$ is the scale factor of the expansion of the universe. 

When assuming Gibbons-Hawking relation for temperature of the horizon, the relative energy fluctuation is proportional to the square of the Hubble parameter, as given in expression (\ref{eqn:flucdof12}). This implies that, during matter dominated era, the relative fluctuations becomes 
$\left| {\sigma_{_H}^2}/{\left<E_H\right>^2}\right| \sim [(\hbar G)/(2\pi)] H_0^2 \Omega_{mo} a^{-3},$ while at the end de Sitter phase,  it takes the form, $\left| {\sigma_{_H}^2}/{\left<E_H\right>^2}\right|  \sim [(\hbar G)/(2\pi)] H_0^2 \Omega_{\Lambda},$ which is a constant. According to this, the relative energy fluctuations of the horizon will decrease as the universe expands and attain a constant value at the end de Sitter epoch.

Now we will analyse the evolution of the relative fluctuations given in equation(\ref{eqn:fluc11}), corresponding to Kodama-Hayward temperature.
For the analysis 
we need to obtain $\dot H$ and $(d\dot H/dH).$ Using equation (\ref{eqn:Hubble1}), one can arrive it,  
\begin{equation}
\dot H = -\frac{3}{2} H_{0}^2 \Omega_{m0} a^{-3}, \quad \quad \frac{d\dot H}{dH} = -3H.
\end{equation}
As a result the relative fluctuations of horizon reduces to, 
\begin{equation}\label{eqn:fluc121}
\frac{\sigma_{_H}^2}{\left<E_{_H}\right>^2}  =\frac{\hbar  G}{2\pi}H^2\frac{\left(-5-3\frac{\dot H}{H^2}\right)}{\left(-1-\frac{\dot H}{ H^2}\right)}. 
\end{equation} 
In the matter dominated era, we have,
\begin{equation}
\frac{\dot H}{H^2} \sim \frac{(-3/2) H_0^2 \Omega_{m0} a^{-3}}{H_0^2 \Omega_{m0}a^{-3}} \sim -\frac{3}{2}.
\end{equation}
As a result, the relative energy fluctuations during the matter dominated era becomes,
\begin{equation}
\label{eqn:flucm1}
\left| \frac{\sigma_{_H}^2}{\left<E_{_H}\right>^2} \right|_m \sim \frac{\hbar G}{2\pi} H_0^2 \Omega_{m0} a^{-3}.
\end{equation}
While, in the future asymptotic limit, the $\Lambda$CDM model tends to the end de Sitter epoch, for which $\dot H \sim 0,$ and the relative fluctuations reduces to,
\begin{equation}\label{eqn:fluc121}
\left| \frac{\sigma_{_H}^2}{\left<E_{_H}\right>^2} \right|_{\Lambda} \to \frac{5\hbar  G}{2\pi }H^2, 
\end{equation}
which is a constant, since in the end epoch $H \sim H_0\sqrt{\Omega_{\Lambda}},$ a constant.  The intriguing fact about this result is that, it is five times larger than the corresponding magnitude of the fluctuations of the horizon energy of the end de Sitter epoch, when one use the Gibbons-Hawking temperature instead of Kodama-Hayward temperature. The evident reason for this is that, the Kodama-Hayward temperature involves the time rate of the Hubble parameter and because of that, the fluctuations  depend on the variation of the time rate of the Hubble parameter with the Hubble parameter itself.  

At the present epoch, the scale factor, $a=a_0=1.$ 
Hence during matter dominated era, the effective vale of the scale factor must be less than one. This implies that, the relative fluctuations of horizon energy during end epoch is less than that in the matter dominated era. Hence we can conclude that, the relative energy fluctuations of the Hubble horizon is decreasing, as the universe expands and attains a constant minimum value at the end de Sitter epoch. It turns out that,  
the magnitude of the relative fluctuations is of the order of $10^{-119},$ when one use the standard values,  $H_0 \sim 69 km/s/Mpc$ and $\Omega_{\Lambda} \sim 0.7$\cite{Planck:2018vyg}.  
Further, it may be noted that, during the matter dominated era, both $\dot H/H^2,$ and $d\dot H/dH,$ contribute, while during the end de Sitter epoch only $d\dot H/dH,$ contributes. This is the primary reason, for the relatively larger strength of the horizon energy fluctuations during matter dominated era.

\section{Degrees of freedom on the horizon and the fluctuations in horizon energy}

In this section, we analyse  
fluctuations of the horizon energy following the second approach, that is by using the degrees of freedom on the horizon.  
For any conventional thermal system, it is possible to express the energy of that system in terms of the degrees of freedom residing in the system. We should note that the horizon of the universe has both entropy and temperature\cite{Hawking:1976de,Bekenstein:1975tw}, and hence it can be treated as a thermodynamic system. Then the energy of the horizon can be written using the equipartition rule as, 
\begin{equation}\label{eqn:EHD}
\left<E_{_H} \right>=\frac{1}{2}N_{_{sur}} k_{_B} T_H,
\end{equation}
where, 
\begin{equation}
\label{eqn:dof}
N_{sur} = \frac{A_H}{L_P^2},
\end{equation}
is the degrees of freedom on the apparent horizon, with
$A_{H}={4\pi} \left(\frac{1}{H}\right)^2,$ the area of the horizon and $L_p$ is Planck length. This notion of the equipartition rule has been widely used in the literature to derive dynamics of the universe from the horizon thermodynamics\cite{padmanabhan2012emergence}.
The essential idea here is that, the Planck area, $L_p^2,$ is taken to be the size of a single degree of freedom on the horizon. This way of defining the degrees of freedom was originally used by Padmanabhan\cite{padmanabhan2012emergence}, in explaining the expansion of the universe as the emergence of cosmic space with the progress of cosmic time.
According to this approach, as the universe approaches the end  de Sitter epoch, the degrees of freedom on the horizon is balanced by that residing within the horizon.
For more details of the various aspects of this approach, see\cite {padmanabhan2012emergence, Krishna:2017vmw}. 

The horizon surface degrees of freedom defined above is fundamentally geometrical, but it acquires gravitational nature through its connection to horizon entropy. In Einstein's gravity the entropy of the horizon is given by Benkestein-Hawking relation, $S=A/4L_P^2.$ Since entropy counts the microscopic degrees of freedom (as $N = 4 S$), the area become directly linked to the gravitational micro states. In this sense, gravity and geometry are deeply connected through the horizon thermodynamics. A more detailed discussion of this connection can be found in reference \cite{Padmanabhan:2009kr}.

We are interested in obtaining the horizon energy fluctuation using the expression for energy in equation(\ref{eqn:EHD}). Using the equations (\ref{eqn:fluc123}) and (\ref{eqn:dof}), we arrive at the  
expression for relative energy fluctuations in terms of the surface degrees of freedom as,
\begin{equation}\label{eqn:flucH1}
\left|\frac{\sigma_{_H}^2}{\left<E_H\right>^2}\right|  = \frac{2}{N_{_{sur}}}\left(\frac{d(\log N_{_{sur}})}{dT}T+1\right).
\end{equation}
As in the previous section, we will now go for obtaining the evolution of energy fluctuations, for the two cases of the horizon temperature, the Gibbons-Hawking temperature and Kodama-Hayward temperature.

\subsection{\textbf{Energy Fluctuations assuming the Gibbons-Hawking and Kodama-Hayward relations for the temperature}}

Let us first go for the fluctuations using the Gibbons-Hawking temperature. Equation (\ref{eqn:flucH1}) can be rewritten as,
\begin{equation}\label{eqn:flucH2}
\left|\frac{\sigma_{_H}^2}{\left<E_H\right>^2}\right|  = \frac{2}{N_{_{sur}}}\left(\frac{d(\log N_{_{sur}})}{dH} \left(\frac{dT}{dH}\right)^{-1} T+1\right).
\end{equation}
On substituting the expressions for $N_{surf}$ and temperature, we found that,
\begin{equation}
\label{eqn:flucdof13}
\left|\frac{\sigma_{_H}^2}{\left<E_H\right>^2}\right|  = \frac{2}{N_{sur}}= \frac{\hbar G}{2\pi} H^2,
\end{equation}
where we have used the fundamental relation for the Planck length, $L_p=\sqrt{\hbar G}.$ The above expression matches exactly with the relative fluctuations of the horizon energy derived in the previous section using the Gibbons–Hawking temperature. What is particularly noteworthy, is the additional insight that the relative energy fluctuations is inversely proportional to the number of degrees of freedom on the horizon. This behaviour mirrors that of an ordinary thermodynamic system, where fluctuations diminish as the number of degrees of freedom increases. In the cosmological context, as the universe expands, the area of the Hubble horizon grows, leading to an increase in the associated degrees of freedom. In the asymptotic limit, the universe approaches a de Sitter phase, where the horizon becomes stationary and the degrees of freedom attains a constant maximum value. Consequently, the relative energy fluctuations steadily decreases and eventually becomes constant in the final de Sitter epoch.

Let us now consider the evolution of the relative energy fluctuations using the Kodama-Hayward temperature.
From equations (\ref{eqn:T3}) and (\ref{eqn:dof}) we can express the relative energy  fluctuations as, 
\begin{equation}\label{fldof}
\frac{\sigma_{_H}^2}{\left<E_H\right>^2}  = \frac{2}{N_{_{sur}}} 
\left(\frac{-2\left(1+\frac{\dot H}{2H^2} \right)}{1-\frac{\dot H}{2H^2}+\frac{1}{2H} \frac{d\dot H}{dH} } + 1 \right).
\end{equation}
Even though the above equation looks different from that obtained using the canonical approach, 
a simple algebraic manipulations will show that, the above equation is identical to equation (\ref{eqn:fluc11}). 
As in the previous case this equation, also depends on the degrees of freedom inversely.  

\subsection{\textbf{Evolution of fluctuations in the context of $\Lambda$CDM model}}

When the Gibbons–Hawking temperature is used, the evolution of the relative fluctuations, in the context of the $\Lambda$CDM model, is always proportional to $H^2.$ As a result, during the matter-dominated era, the fluctuations scale as $\Omega_{m0} a^{-3},$ while in the asymptotic de Sitter epoch, they approach a constant value proportional to $\Omega_{\Lambda}.$

In extracting the evolution of the fluctuations, corresponding to Kodama-Hayward temperature,  in the context of $\Lambda$CDM model, we need to have the expressions for  $d\dot H/dH$ and $\dot H/H^2,$ as given in the previous section. As a result we can rewrite the expression (\ref{fldof}) as,
\begin{equation}
\label{eqn:flucH3}
\frac{\sigma_{_H}^2}{\left<E_H\right>^2} = \frac{2}{N_{sur}} \left(\frac{-2(1 + \frac{\dot H}{2H^2} )}{1 - \frac{\dot H}{2H^2}-\frac{3}{2} } + 1 \right).
\end{equation}
For the matter-dominated era, substituting the expression for ${\dot H}/{H^2},$  we get,
\begin{equation}
\left|\frac{\sigma_{_H}^2}{\left<E_H\right>^2}\right| = \frac{\hbar G}{2\pi} H_0^2 \Omega_{m0} a^{-3}.
\end{equation}
The above equation shows that, irrespective of the difference in the nature of evolution,  the relative energy fluctuations have the same form in the matter dominated era for both the temperatures. 
For end de Sitter epoch, at which $\dot H \sim 0,$ equation (\ref{eqn:flucH3}) takes the form,
\begin{equation}
\label{eqn:dof123}
\left|\frac{\sigma_{_H}^2}{\left<E_H\right>^2}\right| = 5 \left(\frac{2}{N_{sur}}\right) = 5 \left( \frac{\hbar G}{2\pi} \right) H^2.
\end{equation} 
This is also identical to the corresponding relation obtained in the previous section.

\section{Can determination of relative fluctuations resolve the cosmological constant problem?}

In the preceding section, we examined the relative energy fluctuations of the Hubble horizon by considering two distinct definitions of horizon temperature: (i) the Gibbons-Hawking temperature, which depends solely on the Hubble parameter, and (ii) the Kodama-Hayward temperature, which also incorporates the time derivative of the Hubble parameter. The evolution of these fluctuations was analysed using both a canonical ensemble approach and an alternative method based on the horizon degrees of freedom. It was found, using the second approach, that the relative energy fluctuations is inversely proportional to the number of degrees of freedom on the horizon during the end de Sitter epoch, as given in equations (\ref{eqn:flucdof12}) and (\ref{eqn:flucdof13}). Now the horizon degrees of freedom is directly proportional to the area of the horizon (and the thus the entropy of the horizon too\cite{Komatsu:2021ncs}). It is well known that the Hubble horizon area is  proportional to $1/H^2.$ Since the evolution of $H$ is governed by the nature of the cosmic components contained within the horizon, the degrees of freedom on the horizon can be considered to be closely linked to the energy density of these components. Therefore the degrees of freedom on the horizon and hence the fluctuations of horizon energy at the end Sitter epoch corresponds to the late acceleration, are essentially controlled by the cosmological constant, the dark energy density, the dominant component in that epoch.

Let us restrict our analsysis to that based on Gibbons-Hawking temperature for the further analysis in this section. In equation (\ref{eqn:flucdof13}), substitute $\hbar G = L_p^2,$ and rewrite 
the relative fluctuations in the end de Sitter epoch 
as,
\begin{equation}
\left|\frac{\sigma_{_H}^2}{\left<E_H\right>^2}\right|= \left(\frac{L_p^2}{2\pi} \right) H^2.
\end{equation} 
Following the standard Friedmann equation, $H^2=\Lambda/3,$ (where $\Lambda$ is the cosmological constant) corresponding end de Sotter epoch, this equation can be re-expressed as,
\begin{equation}
\left|\frac{\sigma_{_H}^2}{\left<E_H\right>^2}\right| = \frac{L_p^2 \Lambda}{6\pi }
\end{equation}
This gives a simple relationship between the Planck constant and the cosmological parameter in terms of the relative energy fluctuations of the horizon as,
\begin{equation}
L_p^2 \Lambda = 6\pi \left|\frac{\sigma_{_H}^2}{\left<E_H\right>^2}\right|.
\end{equation}
The above equation is a relation which connect two length scales, the cosmological constant, $\Lambda,$ corresponds to the late accelerated epoch of the universe, and the Planck length, which has relevance in the very stage of the universe. If one can able to find the relative energy fluctuations by any independent method, then it gives a hope to constraint the magnitude of the product $L_p^2\Lambda.$ This will in turn gives a valuable insight that could help in addressing the cosmological constant problem.

Let us now try to assess the value of the relative fluctuations of the horizon energy in an approximate way using the magnitude of the horizon entropy. It is well known that, the degrees of freedom on the horizon, is given as $N_{surf}=4 S_H/k_{_B}$ So the relative energy fluctuations at the de Sitter epoch become, $\left|{\sigma_{_H}^2}/{\left<E_H\right>^2}\right| = {1/(2 S_H)}.$ let us consider the entropy of the Hubble horizon during the de Sitter epoch is of the order of $S_H \sim 10^{123}$\cite{Komatsu:2021ncs}. This implies that the product $L_p^2\Lambda$ is of the order of,
\begin{equation}
L_p^2 \Lambda \sim 3\pi \times 10^{-123}.
\end{equation}
Taking the Planck length, $L_P \sim 10^{-35}$m, we assess the order of magnitude of the  cosmological constant as, $\Lambda \sim 10^{-53},$ in its original unit of $[meter]^{-2}.$ This on converting to the conventional units, given the density of dark energy corresponds to cosmological constant as, $\rho_{\Lambda} \sim 10^{-27} kg. m^{-3},$ or $\rho_{\Lambda} \sim 10^{-47} (GeV)^{4}.$ Here an order of calculations gives almost the correct magnitude of the density corresponding to the cosmological constant. Thus an independent determination of the relative fluctuations of energy on the horizon of the late de Sitter epoch, can solve the cosmological constant problem. This method depends on the magnitude of the horizon entropy, upon which we doesn’t have any direct access.

As mentioned above only a direct determination of the horizon fluctuations during the end de Sitter epoch can lead to a possible way out to extract the cosmological constant. Even though there doesn’t exist any direct method to assess the horizon fluctuations, it is possible to find a constraint as described below. The relative energy fluctuations of the horizon are proportional to $H^2.$ As the universe expands, the Hubble parameter steadily decreases and eventually reaches a constant minimum value in the final de Sitter phase. Therefore, the relative energy fluctuations also decrease and finally reaches a constant value at end the de Sitter epoch. Now, the present value of the relative energy fluctuation can be obtained as, 
\begin{equation}
\left|\sigma^2_{_H}/E_{_H}^2\right|_0 \sim \frac{\hbar G H_0^2}{2\pi c^5}.   
\end{equation}
In the literature, there exist two main observational determinations of $H_0$: (1) from the CMBR\cite{Planck:2018vyg} observations and (2) from supernova observational data set\cite{SupernovaCosmologyProject:1996grv}. Let us rewrite the expression for the relative energy fluctuation for convenience as, 
\begin{equation}
\left|\frac{\sigma_H^2}{E_H^2}\right|_0\simeq\frac{H_0^2 t_P^2}{2\pi},
\end{equation}
where, $t_P=\sqrt{\frac{\hbar G}{c^5}}=5.391247\times10^{-44}\,\mathrm{s}$, is the Planck time. Below we obtained the current value of the relative fluctuations in horizon energy, using each of these two observational values of $H_0,$ 
\begin{equation}
\left|\frac{\sigma_H^2}{E_H^2}\right|_0 \simeq 2.18\times10^{-123},
\end{equation}
for the CMBR value of the Hubble parameter, $H_0 = 67 \, \mathrm{km \, s^{-1} Mpc^{-1}},$ and 
\begin{equation}
\left|\frac{\sigma_H^2}{E_H^2}\right|_0 \simeq 2.66\times10^{-123}.
\end{equation}
for the  SNe Ia value of the Hubble parameter, $H_0 = 74\ \mathrm{km \,s^{-1} Mpc^{-1}}.$
Since the relative energy fluctuations of the horizon decrease as the universe expands, its value at the end Sitter de Sitter epoch must be less than its current value. That is,
\begin{equation}
\left|\frac{\sigma_H^{2}}{E_H^{2}}\right|_{\mathrm{end-de}}  <  \left|\frac{\sigma_H^{2}}{E_H^{2}}\right|_{0}
\end{equation}
This can be taken as a constraint on the relative energy energy fluctuations at the end de Sitter epochs, which is obtained using the observational data on $H_0,$ the current value of the Hubble parameter. 
Following equation (32) (which says that the relative energy fluctuation is proportional to $L_P^2 \Lambda$ ), one can arrive at, 
\begin{equation}
\Lambda \leq 
0.83 \times 10^{-54}\,\mathrm{m}^{-2}. \, \, \,\text{for} \, H_0 \, \text{from CMBR} 
\end{equation}
or 
\begin{equation}
\Lambda \leq 
1.02 \times 10^{-54}\,\mathrm{m}^{-2}. \, \, \, \text{for} \, H_0 \, \text{from SNe Ia}. 
\end{equation}
Thus, the constraint on relative energy fluctuations can directly place a bound on the cosmological constant.

\section{Conclusions}

Like black hole horizons, cosmological horizons also exhibit thermodynamic properties. Owing to the intrinsic statistical nature, the thermal energy associated with the horizon, undergoes fluctuations that evolve as the universe expands. In this study, we examine the evolution of the relative energy fluctuations of the Hubble horizon. We derive these fluctuations using two complementary approaches: (i) the canonical ensemble framework, and (ii) a method based on the number of degrees of freedom on the horizon. To incorporate temperature in our analysis, we first employ the Gibbons–Hawking temperature, which parallels black hole thermodynamics. Additionally, we use the temperature relation proposed by Kodama and Hayward, which accounts for the dynamical aspects of the cosmological horizon.

We first analysed the relative fluctuations of the horizon energy following the ensemble approach. The relative fluctuations with Gibbons-Hawking temperature for the horizon follows a simple rule,  $\left| {\sigma_{_H}^2}/{\left<E_H\right>^2}\right| = [(\hbar G)/(2\pi)] H^2.$  Observations indicate that, the Hubble parameter decreases with the expansion of the universe and approaches a minimum constant value at the end de Sitter epoch. Therefore, the relative fluctuations of horizon energy decreases as the universe expands and achieve the minimum constant value at the end de Sitter epoch.

The canonical ensemble analysis with Kodama-Hayward temperature for the horizon, also shows that the relative energy fluctuations of the horizon decreases as the universe expands. However the magnitude of the fluctuation is significantly different compared to that obtained using the Gibbons-Hawking temperature. The main reason for this can be attributed to the dynamical nature of temperature. It turns out that, the fluctuations in this case, gets an additional contribution from the term $d\dot H/dH.$ As a cumulative effect, the value of the fluctuations at the end de Sitter epoch, becomes five times larger than that with the  Gibbons-Hawking temperature. It may be interesting to note that, the result in the reference\cite{Komatsu:2021ncs}, where the calculations are restricted to end de Sitter epoch, magnitude of the relative energy fluctuations, is agreeing only with the asymptotic value obtained with the use of Gibbons-Hawking temperature in the present work. An important point to be noted here is that, during the matter dominated era, the relative energy fluctuations, found using Kodama-Hayward temperature,  
$\left| {\sigma_{_H}^2}/{\left<E_H\right>^2}\right|  
\sim [(\hbar G)/(2\pi)] H_0^2 \Omega_{mo} a^{-3},$ is mimicking the evolutionary status of that which found using Gibbons-Hawking temperature. This is due to the compensatory effect between the terms $\dot H/H^2,$ and $d\dot H/dH.$ 

In our second approach, we evaluate the relative energy fluctuations based on the degrees of freedom associated with the Hubble horizon. Since the horizon possesses thermal energy, it is expected to have underlying microscopic degrees of freedom. As the universe expands, the horizon area increases, leading to a corresponding increase in its degrees of freedom. In conventional thermodynamic systems, relative energy fluctuations decrease as the number of degrees of freedom increases. A similar behaviour is observed in our analysis: for both the Gibbons–Hawking and Kodama-Hayward temperatures, the relative fluctuation is inversely proportional to the degrees of freedom. Consequently, the energy fluctuations per unit energy decrease as the horizon expands. This indicates that the horizon exhibits a behaviour analogous to that of ordinary thermal systems. It is also worth noting that the general evolution — and consequently the asymptotic behaviour — of the relative fluctuations remains consistent in both approaches. However, the results exhibit a strong dependence on the choice of the temperature relation used.

In the final section, we proposed that an alternative determination of horizon energy fluctuations can provide a possible way to address the cosmological constant problem. This idea emerges from the dependence of the relative energy fluctuations to the degrees of freedom on the Hubble horizon. The appearance of the Planck area in the horizon degrees of freedom and the behaviour of the Hubble parameter in the asymptotic de Sitter epoch together lead to a relation of the form, $L_p^2 \Lambda \propto \left| {\sigma_{_H}^2}/{\left<E_H\right>^2}\right|.$
Then the standard value for the cosmological constant ($\Omega \sim 0.7),$ leads to $L_p^2 \Lambda \sim 10^{-123}.$ Therefore, if an independent method is developed to evaluate the ratio $\left| {\sigma_{_H}^2}/{\left<E_H\right>^2}\right|,$ it could offer a novel way to assess the cosmological constant. Using a approximation based on the magnitude of horizon entropy, we obtained a corresponding energy density of the cosmological constant as, $\rho_{\Lambda} \sim 10^{-47} (GeV)^4,$ which aligns with the observed value. 

The horizon energy fluctuations might leave observable imprints on the CMBR anisotropies or on large scale structure. Invoking the holographic principle, one may expect that 
fluctuations on the horizon should, in principle, have its imprint on the
on the cosmic components, like CMBR, residing within the Hubble volume. To examine this possibility, let us consider the magnitude of the relative energy fluctuations of the horizon corresponding to the Hubble parameter, $H_{\rm dec} \sim 10^{-13}\ {\rm s^{-1}},$ during the radiation-matter decoupling epoch of the universe. We obtained it as,
$ \left|\frac{\sigma_H}{E_H}\right|_{dec}
\sim 
\sqrt{\frac{\hbar G H_{\rm dec}^{2}}{2\pi c^{5}}}
\sim 10^{-57}.
$
This value is enormously smaller than the standard fluctuations in the radiation component at the time of decoupling, which of the order $10^{-6}$, which acted as the seeds for cosmic structure formation\cite{Smoot}. Consequently, it seems that imprints of relative energy fluctuations of the horizon on cosmic components like radiation, and its after effect on the structure formation, is comparatively much less compared to standard reasons like quantum fluctuations. 

Finally, we want to compare our work with that in reference 
\cite{Komatsu:2021ncs}, where the author explores the energy fluctuations in a pure (late) de Sitter universe. They arbitrarily connect the square root of the relative energy fluctuations with the ratio of the observed cosmological constant with the energy density of vacuum (the theoretical value). Using this the author made an assessment of the magnitude of the dark energy density. In contrast to this, the present study primarily investigates the evolution of relative thermal energy fluctuations associated with the Hubble horizon in an expanding universe. We establish a possible link between the relative energy fluctuations and the cosmological constant.

\begin{center} 
	\textbf{Acknowledgement}
\end{center}

VSN and KPB are thankful to Cochin University of Science and Technology for financial support for research. Thanks are due to Vishnu A Pai, for the valuable comments on the draft form of the manuscript.



\begin{thebibliography}{10}
	\expandafter\ifx\csname url\endcsname\relax
	\def\url#1{\texttt{#1}}\fi
	\expandafter\ifx\csname urlprefix\endcsname\relax\def\urlprefix{URL }\fi
	\expandafter\ifx\csname href\endcsname\relax
	\def\href#1#2{#2} \def\path#1{#1}\fi
	
	\bibitem{Hawking:1976de}
	S.~W. Hawking, {Black Holes and Thermodynamics}, Phys. Rev. D 13 (1976)
	191--197.
	\newblock \href {https://doi.org/10.1103/PhysRevD.13.191}
	{\path{doi:10.1103/PhysRevD.13.191}}.
	
	\bibitem{Bardeen:1973gs}
	J.~M. Bardeen, B.~Carter, S.~W. Hawking, {The Four laws of black hole
		mechanics}, Commun. Math. Phys. 31 (1973) 161--170.
	\newblock \href {https://doi.org/10.1007/BF01645742}
	{\path{doi:10.1007/BF01645742}}.
	
	\bibitem{Hawking:1975vcx}
	S.~W. Hawking, {Particle Creation by Black Holes}, Commun. Math. Phys. 43
	(1975) 199--220, [Erratum: Commun.Math.Phys. 46, 206 (1976)].
	\newblock \href {https://doi.org/10.1007/BF02345020}
	{\path{doi:10.1007/BF02345020}}.
	
	\bibitem{Gibbons:1977mu}
	G.~W. Gibbons, S.~W. Hawking, {Cosmological Event Horizons, Thermodynamics, and
		Particle Creation}, Phys. Rev. D 15 (1977) 2738--2751.
	\newblock \href {https://doi.org/10.1103/PhysRevD.15.2738}
	{\path{doi:10.1103/PhysRevD.15.2738}}.
	
	\bibitem{Bekenstein:1973ur}
	J.~D. Bekenstein, {Black holes and entropy}, Phys. Rev. D 7 (1973) 2333--2346.
	\newblock \href {https://doi.org/10.1103/PhysRevD.7.2333}
	{\path{doi:10.1103/PhysRevD.7.2333}}.
	
	\bibitem{Bekenstein:1975tw}
	J.~D. Bekenstein, {Statistical Black Hole Thermodynamics}, Phys. Rev. D 12
	(1975) 3077--3085.
	\newblock \href {https://doi.org/10.1103/PhysRevD.12.3077}
	{\path{doi:10.1103/PhysRevD.12.3077}}.
	
	\bibitem{Bousso:1999xy}
	R.~Bousso, {A Covariant entropy conjecture}, JHEP 07 (1999) 004.
	\newblock \href {http://arxiv.org/abs/hep-th/9905177}
	{\path{arXiv:hep-th/9905177}}, \href
	{https://doi.org/10.1088/1126-6708/1999/07/004}
	{\path{doi:10.1088/1126-6708/1999/07/004}}.
	
	\bibitem{Bousso:2002ju}
	R.~Bousso, {The Holographic principle}, Rev. Mod. Phys. 74 (2002) 825--874.
	\newblock \href {http://arxiv.org/abs/hep-th/0203101}
	{\path{arXiv:hep-th/0203101}}, \href
	{https://doi.org/10.1103/RevModPhys.74.825}
	{\path{doi:10.1103/RevModPhys.74.825}}.
	
	\bibitem{Bousso:2004tv}
	R.~Bousso, {Cosmology and the S-matrix}, Phys. Rev. D 71 (2005) 064024.
	\newblock \href {http://arxiv.org/abs/hep-th/0412197}
	{\path{arXiv:hep-th/0412197}}, \href
	{https://doi.org/10.1103/PhysRevD.71.064024}
	{\path{doi:10.1103/PhysRevD.71.064024}}.
	
	\bibitem{Unruh:1976db}
	W.~G. Unruh, {Notes on black hole evaporation}, Phys. Rev. D 14 (1976) 870.
	\newblock \href {https://doi.org/10.1103/PhysRevD.14.870}
	{\path{doi:10.1103/PhysRevD.14.870}}.
	
	\bibitem{Davies:1974th}
	P.~C.~W. Davies, {Scalar particle production in Schwarzschild and Rindler
		metrics}, J. Phys. A 8 (1975) 609--616.
	\newblock \href {https://doi.org/10.1088/0305-4470/8/4/022}
	{\path{doi:10.1088/0305-4470/8/4/022}}.
	
	\bibitem{Biswas:2013lna}
	T.~Biswas, R.~Brandenberger, T.~Koivisto, A.~Mazumdar, {Cosmological
		perturbations from statistical thermal fluctuations}, Phys. Rev. D 88~(2)
	(2013) 023517.
	\newblock \href {http://arxiv.org/abs/1302.6463} {\path{arXiv:1302.6463}},
	\href {https://doi.org/10.1103/PhysRevD.88.023517}
	{\path{doi:10.1103/PhysRevD.88.023517}}.
	
	\bibitem{Magueijo:2002pg}
	J.~Magueijo, L.~Pogosian, {Could thermal fluctuations seed cosmic structure?},
	Phys. Rev. D 67 (2003) 043518.
	\newblock \href {http://arxiv.org/abs/astro-ph/0211337}
	{\path{arXiv:astro-ph/0211337}}, \href
	{https://doi.org/10.1103/PhysRevD.67.043518}
	{\path{doi:10.1103/PhysRevD.67.043518}}.
	
	\bibitem{graef2021constraining}
	L.~Graef, Constraining the spectrum of cosmological perturbations from
	statistical thermal fluctuations, Physics Letters B 819 (2021) 136418.
	
	\bibitem{Magueijo:2007wf}
	J.~Magueijo, P.~Singh, {Thermal fluctuations in loop cosmology}, Phys. Rev. D
	76 (2007) 023510.
	\newblock \href {http://arxiv.org/abs/astro-ph/0703566}
	{\path{arXiv:astro-ph/0703566}}, \href
	{https://doi.org/10.1103/PhysRevD.76.023510}
	{\path{doi:10.1103/PhysRevD.76.023510}}.
	
	\bibitem{Magueijo:2006fu}
	J.~Magueijo, L.~Smolin, C.~R. Contaldi, {Holography and the scale-invariance of
		density fluctuations}, Class. Quant. Grav. 24 (2007) 3691--3700.
	\newblock \href {http://arxiv.org/abs/astro-ph/0611695}
	{\path{arXiv:astro-ph/0611695}}, \href
	{https://doi.org/10.1088/0264-9381/24/14/009}
	{\path{doi:10.1088/0264-9381/24/14/009}}.
	
	\bibitem{Biswas:2014kva}
	T.~Biswas, T.~Koivisto, A.~Mazumdar, {Atick-Witten Hagedorn Conjecture, near
		scale-invariant matter and blue-tilted gravity power spectrum}, JHEP 08
	(2014) 116.
	\newblock \href {http://arxiv.org/abs/1403.7163} {\path{arXiv:1403.7163}},
	\href {https://doi.org/10.1007/JHEP08(2014)116}
	{\path{doi:10.1007/JHEP08(2014)116}}.
	
	\bibitem{Brandenberger:2006vv}
	R.~H. Brandenberger, A.~Nayeri, S.~P. Patil, C.~Vafa, {String gas cosmology and
		structure formation}, Int. J. Mod. Phys. A 22 (2007) 3621--3642.
	\newblock \href {http://arxiv.org/abs/hep-th/0608121}
	{\path{arXiv:hep-th/0608121}}, \href
	{https://doi.org/10.1142/S0217751X07037159}
	{\path{doi:10.1142/S0217751X07037159}}.
	
	\bibitem{brandenberger2009string}
	R.~H. Brandenberger, String gas cosmology, String Cosmology: Modern String
	Theory Concepts from the Big Bang to Cosmic Structure (2009) 193--230\href
	{http://arxiv.org/abs/0808.0746} {\path{arXiv:0808.0746}}.
	
	\bibitem{Li:2015egy}
	C.~Li, {Thermal Fluctuations of Dark Matter in Bouncing Cosmology}, JCAP 09
	(2016) 038.
	\newblock \href {http://arxiv.org/abs/1512.06794} {\path{arXiv:1512.06794}},
	\href {https://doi.org/10.1088/1475-7516/2016/09/038}
	{\path{doi:10.1088/1475-7516/2016/09/038}}.
	
	\bibitem{Cai:2009rd}
	Y.-F. Cai, W.~Xue, R.~Brandenberger, X.-m. Zhang, {Thermal Fluctuations and
		Bouncing Cosmologies}, JCAP 06 (2009) 037.
	\newblock \href {http://arxiv.org/abs/0903.4938} {\path{arXiv:0903.4938}},
	\href {https://doi.org/10.1088/1475-7516/2009/06/037}
	{\path{doi:10.1088/1475-7516/2009/06/037}}.
	
	\bibitem{Peebles:1982ff}
	P.~J.~E. Peebles, {Large scale background temperature and mass fluctuations due
		to scale invariant primeval perturbations}, Astrophys. J. Lett. 263 (1982)
	L1--L5.
	\newblock \href {https://doi.org/10.1086/183911} {\path{doi:10.1086/183911}}.
	
	\bibitem{Peebles:1994xt}
	P.~J.~E. Peebles, {Principles of Physical Cosmology}, Princeton University
	Press, 2020.
	
	\bibitem{rephaeli1986correlated}
	Y.~Rephaeli, W.~C. Saslaw, Correlated statistical fluctuations and galaxy
	formation, Astrophysical Journal, Part 1 (ISSN 0004-637X), vol. 309, Oct. 1,
	1986, p. 13-18. 309 (1986) 13--18.
	
	\bibitem{Mimoso:2018juu}
	J.~P. Mimoso, D.~Pavon, {Fluctuations of the flux of energy on the apparent
		horizon}, Phys. Rev. D 97~(10) (2018) 103537.
	\newblock \href {http://arxiv.org/abs/1805.02894} {\path{arXiv:1805.02894}},
	\href {https://doi.org/10.1103/PhysRevD.97103537}
	{\path{doi:10.1103/PhysRevD.97103537}}.
	
	\bibitem{Komatsu:2021ncs}
	N.~Komatsu, {Energy stored on a cosmological horizon and its thermodynamic
		fluctuations in holographic equipartition law}, Phys. Rev. D 105~(4) (2022)
	043534.
	\newblock \href {http://arxiv.org/abs/2112.06359} {\path{arXiv:2112.06359}},
	\href {https://doi.org/10.1103/PhysRevD.105.043534}
	{\path{doi:10.1103/PhysRevD.105.043534}}.
	
	\bibitem{B:2019oil}
	P.~B. Krishna, T.~K. Mathew, {Does holographic equipartition demand a pure
		cosmological constant?}, Mod. Phys. Lett. A 35~(40) (2020) 2050334.
	\newblock \href {http://arxiv.org/abs/1905.03529} {\path{arXiv:1905.03529}},
	\href {https://doi.org/10.1142/S0217732320503344}
	{\path{doi:10.1142/S0217732320503344}}.
	
	\bibitem{Krishna:2017vmw}
	P.~B. Krishna, T.~K. Mathew, {Holographic equipartition and the maximization of
		entropy}, Phys. Rev. D 96~(6) (2017) 063513.
	\newblock \href {http://arxiv.org/abs/1702.02787} {\path{arXiv:1702.02787}},
	\href {https://doi.org/10.1103/PhysRevD.96.063513}
	{\path{doi:10.1103/PhysRevD.96.063513}}.
	
	\bibitem{Jacobson:1995ab}
	T.~Jacobson, {Thermodynamics of space-time: The Einstein equation of state},
	Phys. Rev. Lett. 75 (1995) 1260--1263.
	\newblock \href {http://arxiv.org/abs/gr-qc/9504004}
	{\path{arXiv:gr-qc/9504004}}, \href
	{https://doi.org/10.1103/PhysRevLett.75.1260}
	{\path{doi:10.1103/PhysRevLett.75.1260}}.
	
	\bibitem{Eling:2006aw}
	C.~Eling, R.~Guedens, T.~Jacobson, {Non-equilibrium thermodynamics of
		spacetime}, Phys. Rev. Lett. 96 (2006) 121301.
	\newblock \href {http://arxiv.org/abs/gr-qc/0602001}
	{\path{arXiv:gr-qc/0602001}}, \href
	{https://doi.org/10.1103/PhysRevLett.96.121301}
	{\path{doi:10.1103/PhysRevLett.96.121301}}.
	
	\bibitem{Padmanabhan:2003pk}
	T.~Padmanabhan, {Gravitational entropy of static space-times and microscopic
		density of states}, Class. Quant. Grav. 21 (2004) 4485--4494.
	\newblock \href {http://arxiv.org/abs/gr-qc/0308070}
	{\path{arXiv:gr-qc/0308070}}, \href
	{https://doi.org/10.1088/0264-9381/21/18/013}
	{\path{doi:10.1088/0264-9381/21/18/013}}.
	
	\bibitem{Padmanabhan:2009kr}
	T.~Padmanabhan, {Equipartition of energy in the horizon degrees of freedom and
		the emergence of gravity}, Mod. Phys. Lett. A 25 (2010) 1129--1136.
	\newblock \href {http://arxiv.org/abs/0912.3165} {\path{arXiv:0912.3165}},
	\href {https://doi.org/10.1142/S021773231003313X}
	{\path{doi:10.1142/S021773231003313X}}.
	
	\bibitem{Padmanabhan:2012bs}
	T.~Padmanabhan, {Equipartition energy, Noether energy and boundary term in
		gravitational action}, Gen. Rel. Grav. 44 (2012) 2681--2686.
	\newblock \href {http://arxiv.org/abs/1205.5683} {\path{arXiv:1205.5683}},
	\href {https://doi.org/10.1007/s10714-012-1418-4}
	{\path{doi:10.1007/s10714-012-1418-4}}.
	
	\bibitem{Cai:2005ra}
	R.-G. Cai, S.~P. Kim, {First law of thermodynamics and Friedmann equations of
		Friedmann-Robertson-Walker universe}, JHEP 02 (2005) 050.
	\newblock \href {http://arxiv.org/abs/hep-th/0501055}
	{\path{arXiv:hep-th/0501055}}, \href
	{https://doi.org/10.1088/1126-6708/2005/02/050}
	{\path{doi:10.1088/1126-6708/2005/02/050}}.
	
	\bibitem{Akbar:2006kj}
	M.~Akbar, R.-G. Cai, {Thermodynamic Behavior of Friedmann Equations at Apparent
		Horizon of FRW Universe}, Phys. Rev. D 75 (2007) 084003.
	\newblock \href {http://arxiv.org/abs/hep-th/0609128}
	{\path{arXiv:hep-th/0609128}}, \href
	{https://doi.org/10.1103/PhysRevD.75.084003}
	{\path{doi:10.1103/PhysRevD.75.084003}}.
	
	\bibitem{Akbar:2006er}
	M.~Akbar, R.-G. Cai, {Friedmann equations of FRW universe in scalar-tensor
		gravity, f(R) gravity and first law of thermodynamics}, Phys. Lett. B 635
	(2006) 7--10.
	\newblock \href {http://arxiv.org/abs/hep-th/0602156}
	{\path{arXiv:hep-th/0602156}}, \href
	{https://doi.org/10.1016/j.physletb.2006.02.035}
	{\path{doi:10.1016/j.physletb.2006.02.035}}.
	
	\bibitem{Hareesh}
	T.~Hareesh, P.~B. Krishna, T.~K. Mathew,
	\href{https://dx.doi.org/10.1088/1475-7516/2019/12/024}{First law of
		thermodynamics and emergence of cosmic space in a non-flat universe}, Journal
	of Cosmology and Astroparticle Physics 2019~(12) (2019) 024.
	\newblock \href {https://doi.org/10.1088/1475-7516/2019/12/024}
	{\path{doi:10.1088/1475-7516/2019/12/024}}.
	\newline\urlprefix\url{https://dx.doi.org/10.1088/1475-7516/2019/12/024}
	
	\bibitem{Mahith}
	M.~Mahith, P.~B. Krishna, T.~K. Mathew,
	\href{https://dx.doi.org/10.1088/1475-7516/2018/12/042}{Expansion law from
		first law of thermodynamics}, Journal of Cosmology and Astroparticle Physics
	2018~(12) (2018) 042.
	\newblock \href {https://doi.org/10.1088/1475-7516/2018/12/042}
	{\path{doi:10.1088/1475-7516/2018/12/042}}.
	\newline\urlprefix\url{https://dx.doi.org/10.1088/1475-7516/2018/12/042}
	
	\bibitem{VT}
	H.~B. V~T, P.~B. Krishna, P.~K~V, T.~K. Mathew,
	\href{https://dx.doi.org/10.1088/1361-6382/ac6a39}{Emergence of space and
		expansion of universe}, Classical and Quantum Gravity 39~(11) (2022) 115012.
	\newblock \href {https://doi.org/10.1088/1361-6382/ac6a39}
	{\path{doi:10.1088/1361-6382/ac6a39}}.
	\newline\urlprefix\url{https://dx.doi.org/10.1088/1361-6382/ac6a39}
	
	\bibitem{PhysRevD.107.063511}
	V.~T. Hassan~Basari, P.~B. Krishna, T.~K. Mathew,
	\href{https://link.aps.org/doi/10.1103/PhysRevD.107.063511}{Unified formalism
		for the law of emergence from the first law of thermodynamics}, Phys. Rev. D
	107 (2023) 063511.
	\newblock \href {https://doi.org/10.1103/PhysRevD.107.063511}
	{\path{doi:10.1103/PhysRevD.107.063511}}.
	\newline\urlprefix\url{https://link.aps.org/doi/10.1103/PhysRevD.107.063511}
	
	\bibitem{Padmanabhan:2004kf}
	T.~Padmanabhan, {Gravity as elasticity of spacetime: A Paradigm to understand
		horizon thermodynamics and cosmological constant}, Int. J. Mod. Phys. D 13
	(2004) 2293--2298.
	\newblock \href {http://arxiv.org/abs/gr-qc/0408051}
	{\path{arXiv:gr-qc/0408051}}, \href
	{https://doi.org/10.1142/S0218271804006358}
	{\path{doi:10.1142/S0218271804006358}}.
	
	\bibitem{Padmanabhan:2008zza}
	T.~Padmanabhan, {Gravity as an emergent phenomenon}, Int. J. Mod. Phys. D 17
	(2008) 591--596.
	\newblock \href {https://doi.org/10.1142/S0218271808012310}
	{\path{doi:10.1142/S0218271808012310}}.
	
	\bibitem{Padmanabhan:2007tm}
	T.~Padmanabhan, {Gravity as an emergent phenomenon: A Conceptual description},
	AIP Conf. Proc. 939~(1) (2007) 114--123.
	\newblock \href {http://arxiv.org/abs/0706.1654} {\path{arXiv:0706.1654}},
	\href {https://doi.org/10.1063/1.2803795} {\path{doi:10.1063/1.2803795}}.
	
	\bibitem{padmanabhan2012emergence}
	T.~Padmanabhan, Emergence and expansion of cosmic space as due to the quest for
	holographic equipartition, arXiv preprint arXiv:1206.4916 (2012).
	
	\bibitem{Wald:1999vt}
	R.~M. Wald, {The thermodynamics of black holes}, Living Rev. Rel. 4 (2001) 6.
	\newblock \href {http://arxiv.org/abs/gr-qc/9912119}
	{\path{arXiv:gr-qc/9912119}}, \href {https://doi.org/10.12942/lrr-2001-6}
	{\path{doi:10.12942/lrr-2001-6}}.
	
	\bibitem{Lyn}
	D.~Lynden-Bell, {Negative specific heat in astronomy, physics and chemistry},
	Physica A 263 (1999) 293--304.
	\newblock \href {http://arxiv.org/abs/cond-mat/9812172}
	{\path{arXiv:cond-mat/9812172}}, \href
	{https://doi.org/10.1016/S0378-4371(98)00518-4}
	{\path{doi:10.1016/S0378-4371(98)00518-4}}.
	
	\bibitem{Thir}
	W.~{Thirring}, {Systems with negative specific heat}, Zeitschrift fur Physik
	235~(4) (1970) 339--352.
	\newblock \href {https://doi.org/10.1007/BF01403177}
	{\path{doi:10.1007/BF01403177}}.
	
	\bibitem{Vel}
	L.~{Velazquez}, {Remarks about the thermodynamics of astrophysical systems in
		mutual interaction and related notions}, Journal of Statistical Mechanics:
	Theory and Experiment 3~(3) (2016) 033105.
	\newblock \href {http://arxiv.org/abs/1603.00044} {\path{arXiv:1603.00044}},
	\href {https://doi.org/10.1088/1742-5468/2016/03/033105}
	{\path{doi:10.1088/1742-5468/2016/03/033105}}.
	
	\bibitem{chatterjee303030black}
	A.~Chatterjee, P.~Majumdar, Black hole entropy: quantum versus thermal
	fluctuations (2003), arXiv preprint gr-qc/0303030.
	
	\bibitem{Artymowski:2018pyg}
	M.~Artymowski, J.~Mielczarek, {Quantum Hubble horizon}, Eur. Phys. J. C 79~(7)
	(2019) 632.
	\newblock \href {http://arxiv.org/abs/1806.03924} {\path{arXiv:1806.03924}},
	\href {https://doi.org/10.1140/epjc/s10052-019-7131-7}
	{\path{doi:10.1140/epjc/s10052-019-7131-7}}.
	
	\bibitem{Landau:1980mil}
	L.~D. Landau, E.~M. Lifshitz, {Statistical Physics, Part 1}, Vol.~5 of Course
	of Theoretical Physics, Butterworth-Heinemann, Oxford, 1980.
	
	\bibitem{Muhsinath:2022cij}
	M.~Muhsinath, H.~Basari V.~T., T.~K. Mathew, {Modified expansion law with
		Kodama\textendash{}Hayward temperature for the horizon}, Gen. Rel. Grav.
	55~(2) (2023) 43.
	\newblock \href {http://arxiv.org/abs/2211.01739} {\path{arXiv:2211.01739}},
	\href {https://doi.org/10.1007/s10714-023-03091-x}
	{\path{doi:10.1007/s10714-023-03091-x}}.
	
	\bibitem{Krishna:2022zzw}
	P.~B. Krishna, H.~B.~V. T., T.~K. Mathew, {Emergence of cosmic space and its
		connection with thermodynamic principles}, Gen. Rel. Grav. 54~(6) (2022) 58.
	\newblock \href {http://arxiv.org/abs/2205.14868} {\path{arXiv:2205.14868}},
	\href {https://doi.org/10.1007/s10714-022-02941-4}
	{\path{doi:10.1007/s10714-022-02941-4}}.
	
	\bibitem{Srednicki:1993im}
	M.~Srednicki, {Entropy and area}, Phys. Rev. Lett. 71 (1993) 666--669.
	\newblock \href {http://arxiv.org/abs/hep-th/9303048}
	{\path{arXiv:hep-th/9303048}}, \href
	{https://doi.org/10.1103/PhysRevLett.71.666}
	{\path{doi:10.1103/PhysRevLett.71.666}}.
	
	\bibitem{easther1999holography}
	R.~Easther, D.~Lowe, Holography, cosmology, and the second law of
	thermodynamics, Physical Review Letters 82~(25) (1999) 4967--4970.
	
	\bibitem{Cai:2008gw}
	R.-G. Cai, L.-M. Cao, Y.-P. Hu, {Hawking Radiation of Apparent Horizon in a FRW
		Universe}, Class. Quant. Grav. 26 (2009) 155018.
	\newblock \href {http://arxiv.org/abs/0809.1554} {\path{arXiv:0809.1554}},
	\href {https://doi.org/10.1088/0264-9381/26/15/155018}
	{\path{doi:10.1088/0264-9381/26/15/155018}}.
	
	\bibitem{hashemi2015hawking}
	M.~Hashemi, S.~Jalalzadeh, S.~Vasheghani~Farahani, Hawking temperature and the
	emergent cosmic space, General Relativity and Gravitation 47 (2015) 1--12.
	
	\bibitem{Sadeghi:2014gza}
	J.~Sadeghi, J.~Naji, H.~Vaez, B.~Khanpour, {Viscous Cosmology and
		Thermodynamics of Apparent Horizon in Modified Friedman-Robertson-Walkers
		Universe}, Int. J. Theor. Phys. 53~(9) (2014) 3089--3094.
	\newblock \href {https://doi.org/10.1007/s10773-014-2104-y}
	{\path{doi:10.1007/s10773-014-2104-y}}.
	
	\bibitem{Banihashemi:2022htw}
	B.~Banihashemi, T.~Jacobson, A.~Svesko, M.~Visser, {The minus sign in the first
		law of de Sitter horizons}, JHEP 01 (2023) 054.
	\newblock \href {http://arxiv.org/abs/2208.11706} {\path{arXiv:2208.11706}},
	\href {https://doi.org/10.1007/JHEP01(2023)054}
	{\path{doi:10.1007/JHEP01(2023)054}}.
	
	\bibitem{Banihashemi:2022jys}
	B.~Banihashemi, T.~Jacobson, {Thermodynamic ensembles with cosmological
		horizons}, JHEP 07 (2022) 042.
	\newblock \href {http://arxiv.org/abs/2204.05324} {\path{arXiv:2204.05324}},
	\href {https://doi.org/10.1007/JHEP07(2022)042}
	{\path{doi:10.1007/JHEP07(2022)042}}.
	
	\bibitem{Mazumdar:2010sa}
	A.~Mazumdar, J.~Rocher, {Particle physics models of inflation and curvaton
		scenarios}, Phys. Rept. 497 (2011) 85--215.
	\newblock \href {http://arxiv.org/abs/1001.0993} {\path{arXiv:1001.0993}},
	\href {https://doi.org/10.1016/j.physrep.2010.08.001}
	{\path{doi:10.1016/j.physrep.2010.08.001}}.
	
	\bibitem{Allahverdi:2010xz}
	R.~Allahverdi, R.~Brandenberger, F.-Y. Cyr-Racine, A.~Mazumdar, {Reheating in
		Inflationary Cosmology: Theory and Applications}, Ann. Rev. Nucl. Part. Sci.
	60 (2010) 27--51.
	\newblock \href {http://arxiv.org/abs/1001.2600} {\path{arXiv:1001.2600}},
	\href {https://doi.org/10.1146/annurev.nucl.012809.104511}
	{\path{doi:10.1146/annurev.nucl.012809.104511}}.
	
	\bibitem{Planck:2018vyg} N. Aghanim et. al, Astron. Astrophys. 641 (2020) A6
	
	\bibitem{SupernovaCosmologyProject:1996grv} S. Perlmutter et. al, Measurements of the cosmological parameters Omega and Lambda from the first seven supernovae at $z \geq 0.35,$ Astrophys. J. 483 (1997) 565 
	
	\bibitem{Smoot} G. F. Smoot et al., Structure in the KOBE differential microwave radiometer first-year maps, Astrophys. J 396 (1992) L1, doi:10.1086/186504
	
	\bibitem{Gob} M. Goberashvili, Fixing cosmological constant on the event horizon, Eur. Phys. J C 82 (2022) 1049, doi:10.1140/epjc/s10052-022-11033-1
\end{thebibliography}

\end{document}